\documentclass[%
 reprint,
superscriptaddress,
amsmath,amssymb,
aps,
pra,
]{revtex4-2}
\usepackage[switch]{lineno}
\usepackage{graphicx}
\usepackage{dcolumn}
\usepackage{bm}

\usepackage{tabularx}
\usepackage{booktabs} 
\usepackage[detect-all]{siunitx}    
\usepackage{array}      
\usepackage{cellspace} 
\usepackage{ulem}

\begin{document}

\preprint{APS/123-QED}

\title{A Compact Dual-Beam Zeeman Slower for \\ High-Flux Cold Atoms}
\author{Chen Chen}
\affiliation{Advanced Materials, Hong Kong University of Science and Technology (Guangzhou), Guangzhou 511453, China}
\author{Kejun Liu}
\affiliation{Advanced Materials, Hong Kong University of Science and Technology (Guangzhou), Guangzhou 511453, China}
\author{Dezhou Deng}
\affiliation{Advanced Materials, Hong Kong University of Science and Technology (Guangzhou), Guangzhou 511453, China}
\author{Shuchang Ma}
\affiliation{Advanced Materials, Hong Kong University of Science and Technology (Guangzhou), Guangzhou 511453, China}
\author{Peng Zhu}
\affiliation{Advanced Materials, Hong Kong University of Science and Technology (Guangzhou), Guangzhou 511453, China}
\author{Zhichang He}
\affiliation{Advanced Materials, Hong Kong University of Science and Technology (Guangzhou), Guangzhou 511453, China}
\author{J. F. Chen}
\affiliation{Department of Physics, Southern University of Science and Technology, Shenzhen 518055, China}
\affiliation{International Shenzhen Quantum Academy, Shenzhen 518048, China}
\author{Xiaoxiao Wu}
\affiliation{Advanced Materials, Hong Kong University of Science and Technology (Guangzhou), Guangzhou 511453, China}
\author{Peng Chen}
\email{pengchen@hkust-gz.edu.cn}
\affiliation{Advanced Materials, Hong Kong University of Science and Technology (Guangzhou), Guangzhou 511453, China}
\affiliation{Quantum Science Center of Guangdong-Hong Kong-Macao Greater Bay Area, Shenzhen 518045, China}

\date{\today}

\begin{abstract}
We present a compact design of  dual-beam Zeeman slower optimized for efficient production of cold atom applications. Traditional single-beam configurations face challenges from substantial residual atomic flux impacting downstream optical windows, resulting in increased system size, atomic deposition contamination, and a reduced operational lifetime. Our approach employs two oblique laser beams and a capillary-array collimation system to address these challenges while maintaining efficient deceleration. For rubidium ($^{87}$Rb), simulations demonstrate a significant increase in the fraction of atoms captured by a two-dimensional magneto-optical trap (2D-MOT) and nearly eliminate atom-induced contamination probability at optical windows, all within a compact Zeeman slower length of $\sim44$~cm. Experimental validation with Rb and Yb demonstrates highly efficient atomic loading within the same compact design. This advancement represents a substantial improvement for high-flux cold atom applications, providing reliable performance for high-precision metrology, quantum computation and simulation.

\end{abstract}

\maketitle


\section{Introduction}

Cold atoms have become indispensable tools in precision metrology, atomic interferometry, and quantum sensing, due to their unique properties and high sensitivity to external fields~\cite{RevModPhys.87.637,NatureCom.7896,PhysRevLett.133.023401}. Among the various techniques for producing cold-atom ensembles, Zeeman slowers have emerged as a robust and efficient method for decelerating thermal atomic beams to velocities suitable for capture by magneto-optical traps (MOTs). Since their introduction in 1982~\cite{PhysRevLett.48.596}, Zeeman slowers have been widely adopted for species with high thermal velocities, from various types of alkaline, alkaline-earth atoms and magnetic atoms, enabling their use in applications ranging from atomic clocks to Bose-Einstein condensation (BEC)~\cite{PhysRevLett.75.1687,Science.269.5221.198,Natures41586,10.1063/5.0162128}. Most recent advances in neutral atom quantum computation stimulate interest for efficient setup for cooling of different species of atoms~\cite{2023Nature_Rb,2023Nature_Yb,2022NC_Sr,PRXQuantum_Sr}. Especially, an increasing number of proposals suggest that hybrid quantum computation architectures utilizing dual species offer significant advantages ~\cite{PhysRevX_dual_species,PhysRevLett_2022dual_zhan,NP_2024dual,Xuwenchao_2025dual}, and call up the need for more efficient and compact Zeeman slower designs to meet the growing demand for portable and scalable quantum technologies~\cite{chen2022continuous,PhysRevA.111.033314,evered2023high}.\par

Despite the success of traditional Zeeman slowers, their single-beam configuration presents considerable challenges for high-flux cold atoms applications~\cite{PhysRevA.102.013319}. A substantial fraction of uncooled atoms continues to propagate downstream, causing contamination and damage to optical windows or mirrors~\cite{Sci_Rep.s41598, PhysRevA.79.042905}. This issue is particularly acute for species requiring high-temperature ovens such as strontium or ytterbium, where atomic velocities can exceed 500 m/s. It is also a concern for highly reactive atoms, like potassium, which can lead to vacuum leaks at glass-to-metal seals. Mitigating this issue typically requires extending the vacuum chamber lengths and adding protective structures, which increases system complexity and reduces compactness~\cite{10.1063/1.2163977, HOSOYA2023129048}.

In this work, we for the first time present a novel dual-beam Zeeman slower design that addresses these challenges while maintaining high deceleration efficiency. By employing two symmetrically aligned laser beams at small angles relative to the atomic flux and incorporating a capillary-array collimation system, our approach greatly reduces harmful residual atomic flux while achieving comparable or better performance than conventional systems within a compact size~\cite{PhysRevA.96.053415,10.1063/1.3600897,10.1063/1.4808375}. Numerical simulations based on Monte Carlo trajectory analysis demonstrate that this design can increase the fraction of atoms captured by a 2D-MOT by more than two orders of magnitude, while nearly eliminating atom-induced contamination at the optical windows. Experimental validation using Rb and Yb atoms in the same compact platform, demonstrates atomic loading rates of $1.2\times 10^9$ atoms/s for Rb and $8.0\times 10^{10}$ atoms/s for Yb at a background pressure below $10^{-9}$~mbar. This advancement indicates the development of a compact quantum system that utilizes cold atoms from multiple species, establishes a scalable pathway for high-fidelity neutral-atom quantum processors and emergent quantum simulation architectures.

The remainder of this paper is organized as follows: Section II provides detailed descriptions of the experimental setup and theoretical analysis, including the capillary array design and magnetic field configuration. Section III presents simulation results and experimental validation, focusing on deceleration efficiency and harmful flux reduction. Section IV discusses the implications of our findings for various atomic species and applications, while Section V concludes with a summary of key innovations and their broader impact on cold atom research.

\section{Experimental Design and Theoretical Analysis}
\subsection{Experimental Design}
In a Zeeman slower, atoms experience deceleration and Doppler cooling via radiation pressure when scattering photons from a laser beam traveling opposite to the atomic flux. While maximal velocity reduction occurs when this beam directly opposes atomic motion, residual uncooled atoms under such alignment pose collision risks to downstream optical window. By replacing conventional single-beam configurations with dual-angularly laser beams, both protective shielding against residual atomic flux and sustained slowing efficacy are achieved, simultaneously enabling spatial miniaturization of the apparatus.

\begin{figure}
    \centering
    \includegraphics[width=\linewidth]{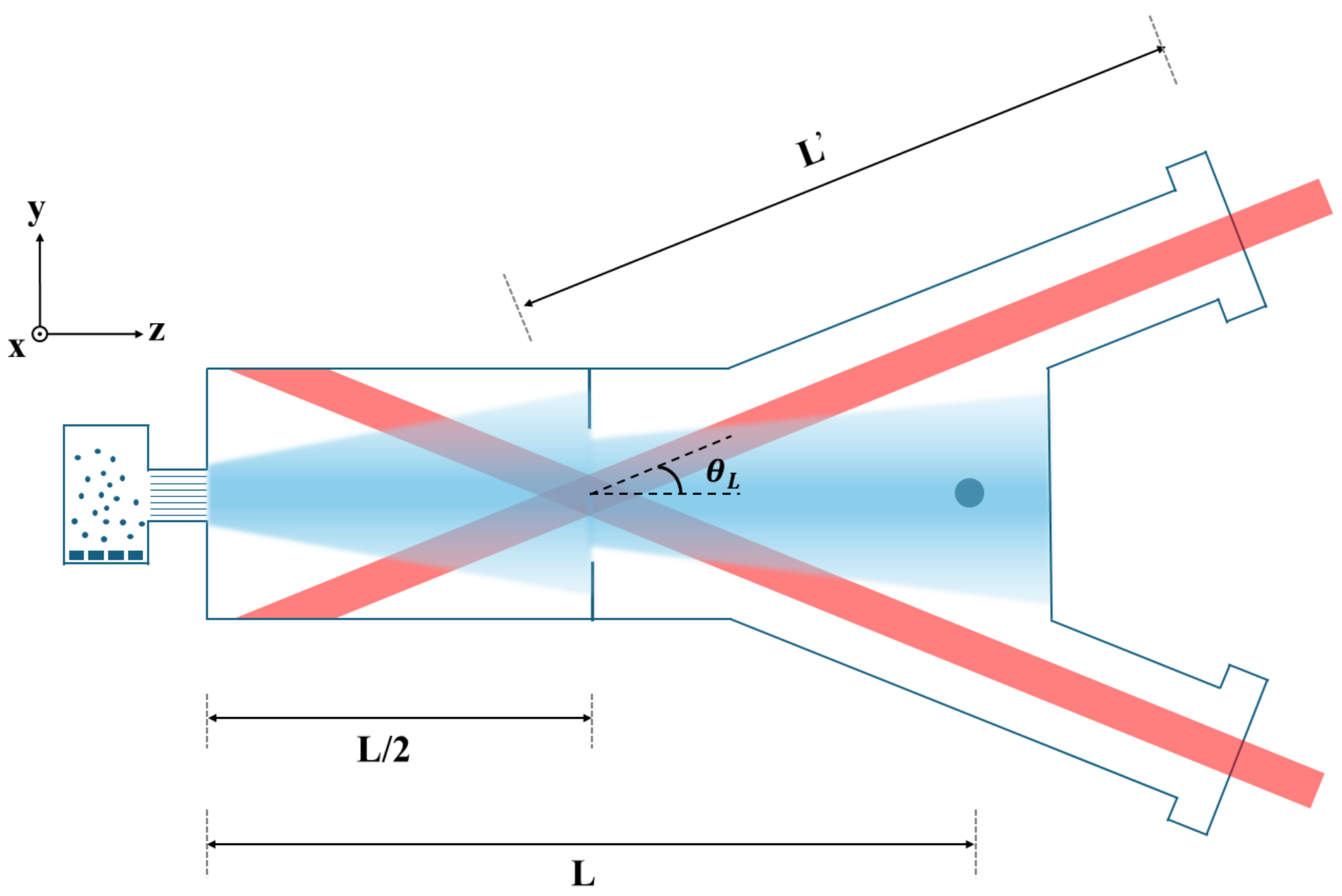}
    \caption{Schematic of the dual-beam small-angle Zeeman slower and capillary array. The atomic beam is effusively emitted from an oven and collimated through a capillary array before entering the Zeeman slower region. The parameter $L$ is the length of the deceleration region, $L{'}$  the length of the Zeeman deceleration light path, and $\theta_L$ the angle between the Zeeman slower beam and the central axis. The blue spot is the center of 2D-MOT, which is positioned immediately after the deceleration region of the Zeeman slower for simplicity.}
    \label{fig1}
\end{figure}

As illustrated in Fig.~\ref{fig1}, atoms effuse from a temperature-stabilized oven and are collimated through a capillary tube array. Two symmetric counter-propagating laser beams are angled at $\theta_L$ with respect to the direction opposing the atomic beam.  Here, $L$ is the deceleration region length and $L'$ specifies the slower beam path length. 

Using a highly directional capillary collimation system reduces atomic scattering toward lateral observation windows during deceleration, greatly improving upon conventional uncollimated configurations where diffuse atomic fluxes compromise vacuum integrity or measurement fidelity. A circle baffle is placed at just the front of the position where two laser beams shoot into the Zeeman-slowing tube, collimating the atomic flux again and reducing the harmful flux. For simplicity in simulation and discussion, the 2D-MOT is placed immediately downstream of the Zeeman slower. This design achieves comparable performance through efficient momentum transfer from the two symmetric counter-propagating beams.

\subsection{Capillary Array and Angular Flux Distribution}\label{Transparent channel}

The capillary array effectively collimates the atomic flux, thereby reducing its angular spread and minimizing the amount of residual flux that reaches downstream optical windows. We analyze the collimation effect in a capillary array composed of $N$ identical cylindrical channels, each with length $l$ and radius $a$. The model assumes diffuse scattering at the channel walls but neglects collisions for atoms propagating along the central axis. As a result, the angular distribution of the effusive atomic flux – the rate at which atoms emerge from the capillary array – is governed by wall collisions. Under equilibrium conditions without macroscopic flow, atomic motion becomes isotropically distributed. The collision rate $v_{\text{col}} $ is related to the atomic number density $ n $ and mean velocity $ \bar{v} $ by $v_{\text{col}} = \frac{1}{4} n \bar{v}$. For long capillary tubes, the axial variation of $ v_{\text{col}}(z) $ is approximated as~\cite{PhysRevApplied.13.014013,10.1063/1.321845}
\begin{equation} 
\frac{v_{\text{col}}(z)}{v_{\text{col}}(0)} = \frac{(1 - W)z}{l} + \frac{W}{2},
\end{equation}
where $ l $ is the tube length and $ z $ represents the position along the capillary, and $ v_{\text{col}}(0)$ is the collision rate at the entrance of the capillary.  $ W $ is the Clausing factor, defined as
\begin{equation} 
W = \left( \frac{8a}{3l} \right) \left( 1 + \frac{8a}{3l} \right)^{-1},
\end{equation} 
arising from the finite size of the channel compared to the atomic mean free path. The angular distribution function $ f(\theta) $ is expressed differently depending on the angle.

For $ 0 < \theta < \arctan(2a/l) $
  
  \begin{equation} \label{eq3}
  f(\theta) \approx \frac{1}{W} - \frac{8}{3\pi}\frac{\theta}{W^2},
  \end{equation} 

For $ \theta > \arctan(2a/l) $

  \begin{equation} \label{eq4}
  f(\theta) \approx \frac{1}{\pi} (1 - W) \frac{\cos^2 \theta}{\sin\theta} + \frac{\cos \theta }{2}.
  \end{equation}
The function $ f(\theta) $ satisfies the normalization condition $ \int f(\theta) d^2 \Omega = \pi $. From Eq.~\ref{eq3}, it is readily found that when $ l \gg a $, the fraction of the collimated atomic flux is limited to $ 0.75/\pi \approx 0.24 $. 
This is consistent with Eq.~\ref{eq4}, which indicates that when $W \approx 0$, $ f(\theta) $ becomes independent of W in this region and integrates to a residual atomic flux fraction of $ 1-0.75/\pi \approx 0.76 $.

The total transmitted atomic flux is given by
\begin{equation} \label{eq5}
\Phi_{\text{tol}} = \frac{\pi}{4} WN n \bar{v} a^2,
\end{equation} 
indicating that the transmission is reduced by the Clausing factor $W$ as a trade-off for the collimation effect provided by the capillary array. 

Therefore, while capillary arrays can effectively direct atomic flux, they do not achieve strong collimation, resulting in a reduction in overall transmission.

\subsection{Magnetic Field and Doppler Deceleration}
The collimated atomic beam produced by the capillary array is then subjected to subsequent slowing using a Zeeman slower. The capture velocity of a Zeeman slower is determined by a carefully designed magnetic field gradient. Under optimal conditions, this spatially varying magnetic field facilitates continuous resonant photon scattering for atoms within the capture velocity range, resulting in maximum radiation pressure that decelerates them to the desired terminal velocity.

The on-resonance deceleration for two laser beams configuration is expressed as
\begin{equation}
    a_{\text{cap}}(z) = -\varepsilon \frac{\hbar k \Gamma}{m} \frac{S(z)}{1 + 2S(z)},
\end{equation}
where $k$ is the wave vector, $\hbar$ is the reduced Planck constant, $\Gamma$ is the natural linewidth, $\varepsilon$ is an empirical deceleration stability coefficient, m is the atomic mass, and $S(z)$ is the spatially dependent saturation parameter of single beam~\cite{Bober2010DesigningZS}.  The capture velocity can be derived using
\begin{equation}
    v(L) = \sqrt{ v_{\text{cap}}^2 - \int_0^L 2 a_{\text{cap}}(z) \, dz },
\end{equation}
where $L$ is the length of deceleration region (see Fig.~\ref{fig1}).  $v(z)$ represents the velocity of an atom at position $z$. The magnetic field $B(z)$, which incorporates the Zeeman shift, is given as~\cite{Bober2010DesigningZS}
\begin{equation}\label{eq8}
    B(z) = \frac{\hbar}{\mu_{\text{eff}}} \left( -\delta_0 - k v(z) \cos\theta_L + \frac{\Gamma\sqrt{ [1 + 2 S(z)] \frac{1 - \varepsilon}{\varepsilon} }}{2}  \right),
\end{equation}
where $\mu_{\text{eff}}B/\hbar$ is the effective magnetic shift of the transition frequency caused by the field intensity $B$. $\delta_0=-kv_{cap}\cos \theta_L$ is the Doppler detuning of the laser frequency from the atomic transition frequency.

For atoms with velocities within the capture velocity, the deceleration is given by
\begin{equation}\label{eq9}
    a(z) = -\frac{\hbar k \Gamma}{m} \frac{S(z)}{1 + 2S(z) +  \frac{4\left( -\delta_0 - k v(z) \cos\theta_L - \frac{\mu_{\text{eff}}}{\hbar} B(z) \right)^2}{\Gamma^2} }.
\end{equation}

\section{Result}
\begin{figure}
    \centering
    \includegraphics[width=0.7\linewidth]{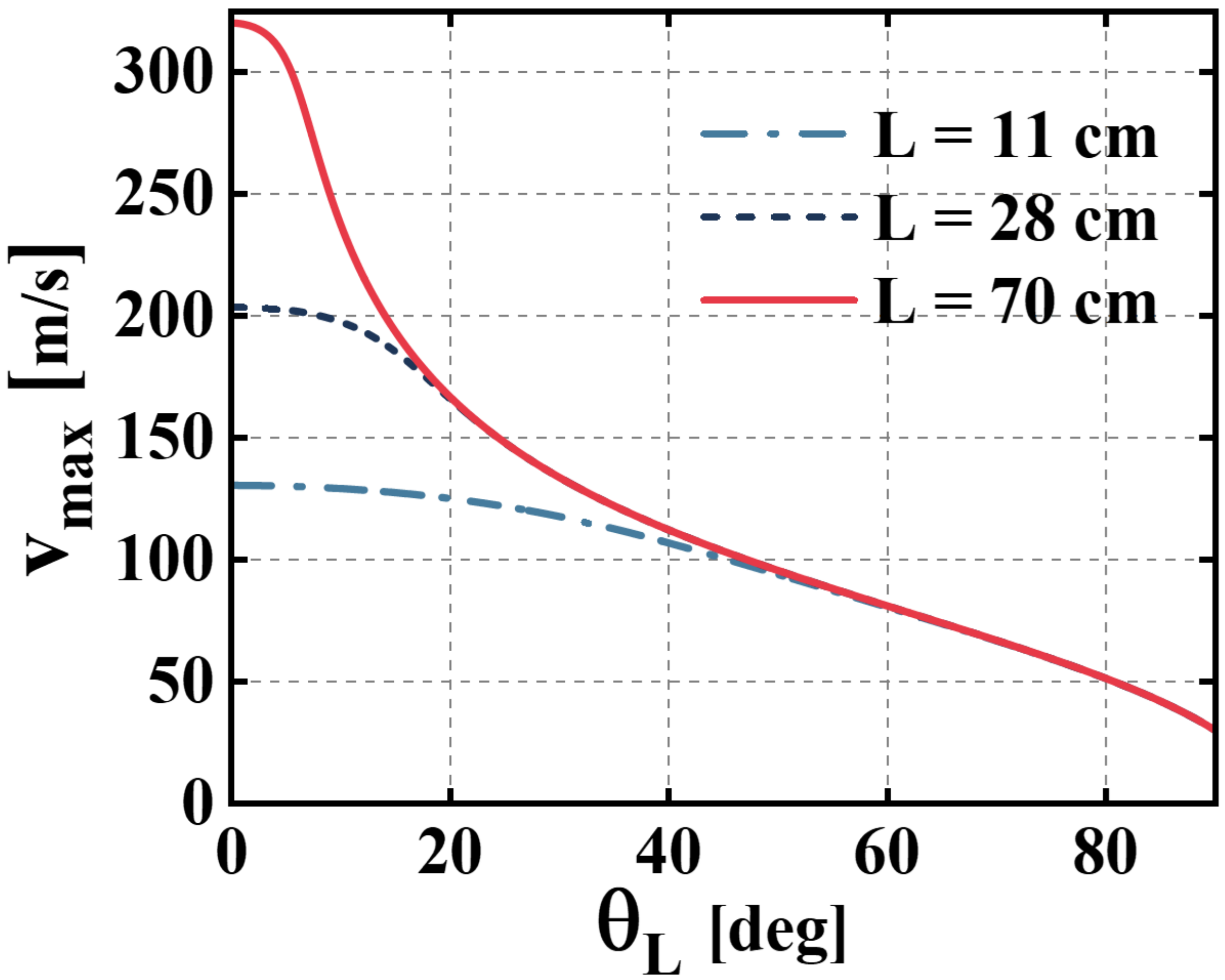}
    \caption{Maximum achievable deceleration velocity of rubidium atoms with different beam angles in the double-beam Zeeman slower. The solid red line, dark blue dotted line, and light blue dotted line correspond to $L=11,28,70$~cm respectively.}
    \label{fig2}
\end{figure}

 We conduct a comprehensive evaluation of the performance of a dual-beam small-angle Zeeman slower using numerical simulations and experimental measurements. Our investigation focuses on three main aspects: (1) determining the range of atomic velocities that can be effectively decelerated under different geometrical configurations, (2) quantifying the reduction of residual atomic flux reaching the vacuum optical window in comparison with a conventional single-beam design, and (3) characterizing the overall deceleration performance of the Zeeman slower. To validate our findings, we experimentally load $^{87}\text{Rb}$ and $^{174}\text{Yb}$ atoms into a 2D-MOT and assess the performance of the Zeeman slower using a fluorescence imaging technique.

\subsection{Maximal Achievable Deceleration Velocity}

We analyze the relationship between the maximal achievable deceleration velocity $v_{cap}$ and the cooling beam angle $\theta_L$, as shown in Fig.~\ref{fig2}. Three configurations of $L$ are examined: $70$~cm (red curve), $28$~cm (blue dashed curve), and $10$~cm (light blue dash-dotted curve). For all configurations, we assume ideal magnetic field profiles (Eq.~\ref{eq8}) to maintain resonance during deceleration. The length $L{'}$ is set to $50$~cm, with the cooling beam diameter fixed at $30$~mm. 

The results demonstrate a strong dependence on the effective deceleration length: longer systems exhibit superior deceleration capabilities comparable to conventional Zeeman slowers, while compact configurations facilitate larger beam angles. 

For the longest configuration ($L = 70$~cm), peak performance reaches $v_{\text{cap}} = 320$ m/s at $\theta_L = 0^\circ$. This system maintain velocities above $200$~m/s until $\theta_L = 14^\circ$, decreasing gradually to $120$~m/s by $\theta_L = 36^\circ$. The velocity response remains relatively flat up to $\theta_L = 5^\circ$, demonstrating that small cooling beam angles could achieve deceleration performance comparable to conventional Zeeman slowers ($\theta_L=0^\circ$).

The intermediate configuration ($L= 28$~cm) shows gradual $v_{\text{max}}$ reduction from an initial 204 m/s at $\theta_L = 0^\circ$, retaining $51$~\% of its peak velocity (104 m/s) even by $\theta_L = 45^\circ$. 

The most compact configuration ($L = 10$~cm) achieve the lowest initial $v_{\text{cap}}=130$~m/s) but exhibit the slowest angular-dependent degradation, suggesting reduced sensitivity to the cooling beam angle.

\begin{figure}
    \centering
    \includegraphics[width=\linewidth]{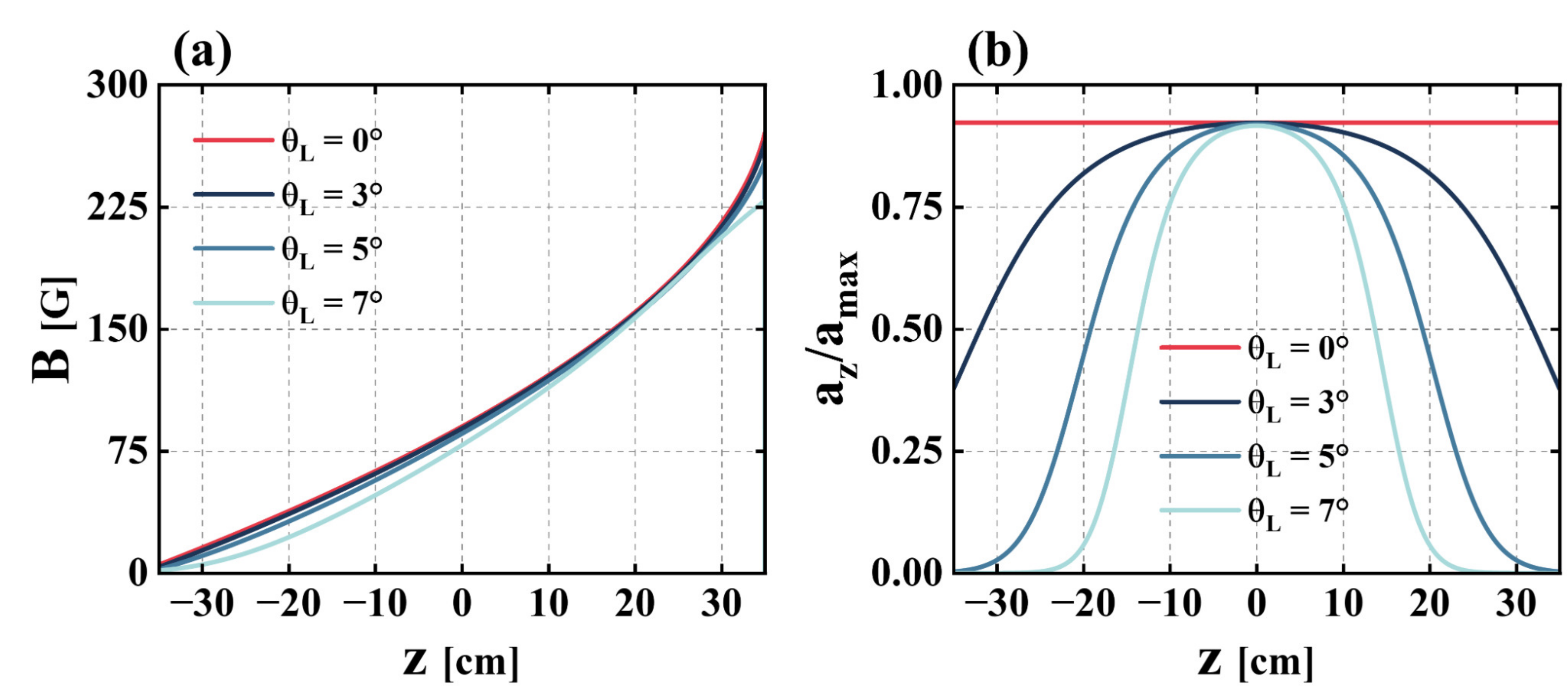}
    \caption{(a) Magnetic field distribution along the Zeeman slower axis for different laser angles $\theta_L$. The deviation from the zero-degree condition is marginal at small angles ($\theta_L \leq 5^\circ$), ensuring effective deceleration.  (b) Doppler deceleration profiles showing reduced efficiency in larger $\theta_L$, with only the central region maintaining optimal acceleration.}
    \label{fig3}
\end{figure}

To assess whether the magnetic field remains comparable to the zero-degree case and is practically feasible under various angles, we simulate the magnetic field distribution along the Zeeman slower axis (Fig.~\ref{fig3}(a)). The results indicate a slight deviation from the zero-degree condition for $\theta_L \leq 7^\circ$, ensuring effective deceleration. These findings align with Eq.~\ref{eq8} and confirm the calculated trends revealed in Fig.~\ref{fig2}. However, as $\theta_L$ increases, the deceleration profiles (Fig.~\ref{fig3}(b)) reveal that the effective deceleration region narrows, with only the central area maintaining maximum acceleration.

\subsection{Harmful Atomic Flux Reduction}
\begin{figure}[htbp]
    \centering
    \includegraphics[width=\linewidth]{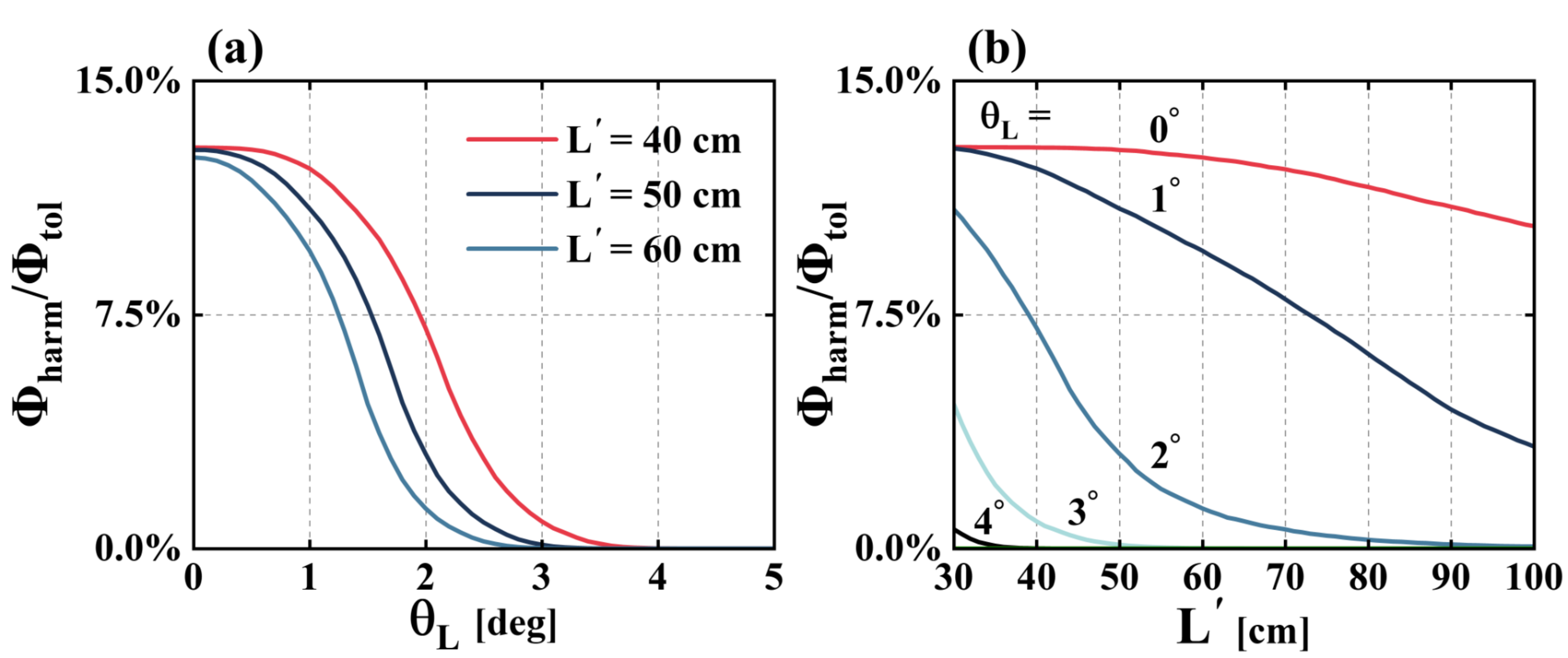}
    \caption{Ratio of harmful atomic flux ($\Phi_{\text{harm}}/\Phi_{\text{tol}}$) as a function of beam angle $\theta_L$ for different $L^{'}$ values. $L$ is set to $70$~cm. (a) The three curves represent $L^{'}=40$~cm, $50$~cm, and $60$~cm from right to left. (b) For each angle $\theta_L$, ratio of harmful atomic flux ($\Phi_{\text{harm}}/\Phi_{\text{tol}}$) is shown as a function of $L^{'}$.  }
    \label{fig4}
\end{figure}

In comparison to a conventional Zeeman slower, our double-beam design greatly decreases the residual atomic flux that reaches the optical windows, which we refer to as harmful flux. We quantify this improvement by comparing the ratio of harmful atomic flux ($\Phi_{\text{harm}}/\Phi_{\text{tol}}$) across different configurations.

For $\theta_L = 0^\circ$, variations in $L{'}$ have minimal impact on the harmful flux ratio (Fig.~\ref{fig4}(a)). However, as shown in Fig.~\ref{fig4}(b), increasing $\theta_L$ introduces significant changes. For example, when $L{'} = 60$~cm, $\Phi_{\text{harm}}/\Phi_{\text{tol}}$ approaches zero at $\theta_L=3^\circ$, indicating that the atomic beam no longer directly strikes the optical window. This behavior highlights the geometric advantages of our compact design.

From Fig.~\ref{fig4}(b), we find that for $L{'}=30$ cm, maintaining the same harmful flux ratio at $\theta_L=2^\circ$ as at $\theta_L=0^\circ$ would require extending the length of the system by $70$~cm. For larger angles, this necessary extension exceeds $150$~cm. Consequently, our double-beam Zeeman slower provides a more compact solution compared to conventional single-beam systems.

\subsection{Overall Deceleration Performance and \\Experimental Validation}

\begin{figure}
    \centering
    \includegraphics[width=\linewidth]{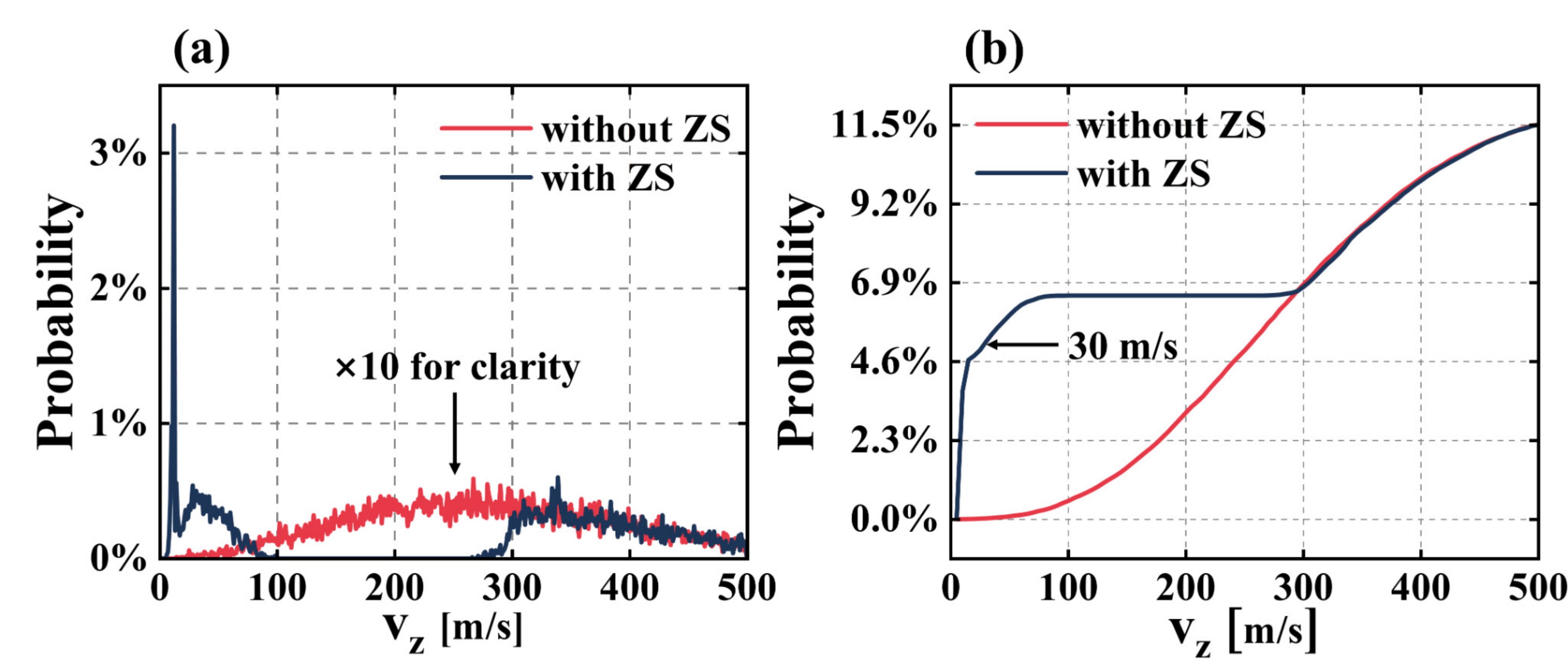}
    \caption{Histogram showing the velocity distribution of atoms after deceleration, with a peak at $12.5\pm0.5$~m/s. The probability without Zeeman slower (red curve) is scaled by $\times$10 for visibility. (b) Cumulative distribution confirms that about $5.2$\% of atoms reach the MOT with velocities between $0-30$~m/s, representing a 235-fold improvement over the atomic beam without Zeeman slower cooling.}
    \label{fig5}
\end{figure}

Figure~\ref{fig5} presents Monte Carlo simulation results for overall deceleration performance under the parameters $L = 70$ cm, $L{'} = 50$ cm, and $\theta_L = 3^\circ$. Spatial variations across the slowing region (spanning coordinates $(x,y)$) influence the effective magnetic field distribution through a position-dependent detuning parameter $S(z)$.  The initial laser detuning is set using $\delta_0 = -kv_{\text{set}} \cos{\theta_L} - \frac{\mu_{\text{eff}} B_0}{\hbar}$, where $B_0$ denotes the magnetic field strength at $z=0$ and $v_{\text{set}} = 300$ m/s. The value of $v_{\text{set}}$ is set to slightly undershoot the maximal achievable deceleration velocity $v_{\text{cap}} = 316$~m/s, ensuring that most atoms fall within the target velocity range of $0$-$30$~m/s.

As shown in Fig.~\ref{fig5}(a), the simulation results demonstrate pronounced deceleration behavior. Atoms with initial velocities below $300$ m/s are predominantly reduced to $0-30$ m/s, while faster atoms exhibit typical Maxwell-Boltzmann distributions. A distinct peak centered at $12.5 \pm 0.5$ m/s is observed, accounting for approximately $3.2\%$ of the atoms. A small fraction of atoms remains in the intermediate velocity range of between $30$-$100$~m/s likely due to inadequate cooling. The cumulative distribution shown in Fig.~\ref{fig5}(b) clearly demonstrates that approximately $5.2\%$ of the atoms enter the region with velocities below $30$~m/s, representing a 235-fold enhancement compared to the atomic beam without Zeeman slower cooling. This result indicates that $75\%$ of the atoms initially below $300$~m/s have been effectively cooled to the targeted region.

Beam blooming effect is a pronounced challenge in Zeeman slower applications, as extensively investigated in the literature~\cite{PhysRevA.81.043424,PhysRevLett.100.053201,10.1063/1.4900577}. This phenomenon originates from the inherent trade-off between longitudinal deceleration and transverse heating: while longitudinal velocities are reduced, the transverse velocity distribution widens, potentially causing unacceptable atomic beam divergence and reducing MOT loading efficiency. Several approaches have been developed to address this challenge, including the incorporation of transverse cooling within the Zeeman slower structure~\cite{PhysRevA.81.043424}, minimizing the distance between the slower exit and MOT region~\cite{PhysRevLett.100.053201}, and ensuring final deceleration occurs within the MOT volume~\cite{10.1063/1.4900577}.

\begin{figure}
    \centering
    \includegraphics[width=0.7\linewidth]{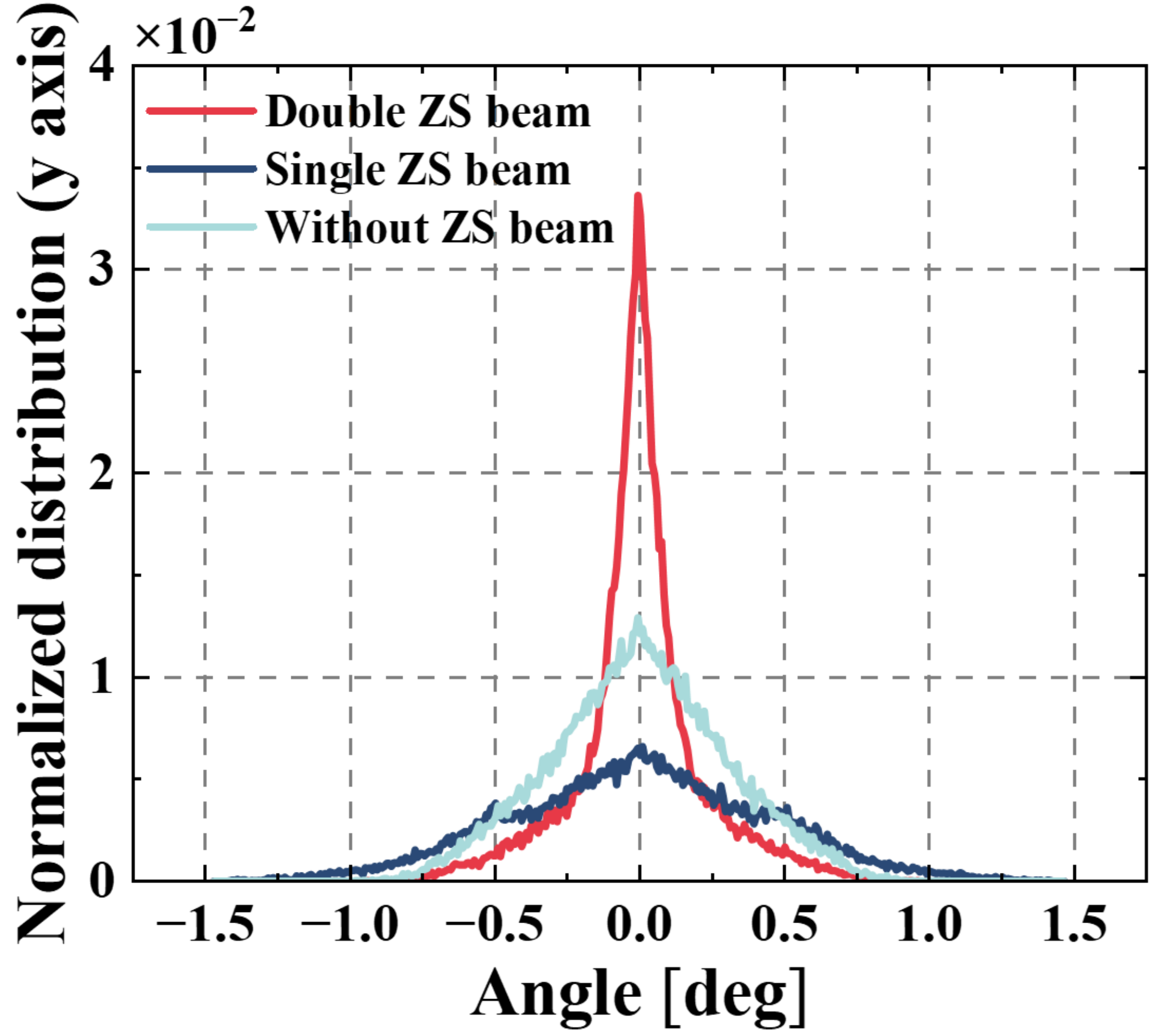}
    \caption{The angular distribution of the atomic cloud along the $y$-axis at the location of the 2D-MOT.}
    \label{fig6}
\end{figure}

 We emphasize the advantages of a dual-beam architecture, where the symmetric configuration induces a self-converging behavior in the atomic trajectory. The dual-beam system exploits two counter-propagating beams to provide an additional transverse cooling effect, effectively reducing beam blooming. As shown in Fig.~\ref{fig6}, this novel approach achieves a dramatic improvement in atomic beam divergence in y direction (the cross-direction of two Zeeman slower beams, see Fig.~\ref{fig1}): while single-beam systems exhibit an beam divergence of $0.32^\circ$ (FWHM), the dual-beam implementation reduces this to $0.08^\circ$ (FWHM). This anti-blooming effect highlights the potential of the dual-beam design for advanced atomic manipulation, particularly for compact applications involving a relatively small cooling beam size in a MOT.

To validate these simulation results and demonstrate the universality of our method, we construct two identical sets of small-scale Zeeman slowers integrated with a shared 2D-MOT for both $^{87}$Rb and $^{174}$Yb in a compact vacuum system with the following specific parameters (powered by Limited Vacuum company): 
\begin{itemize}
    \item Vacuum setup and laser beams: $L = 11.1$~cm, $L{'} = 15$~cm, $\theta_L = 7^\circ$. Zeeman slower beam diameter $w_{\text{ZS}} = 10$~mm (CF16 tubes). The Zeeman slowers for Rb and Yb are oriented perpendicularly to each other, both connected to the common 2D-MOT setup. The 2D-MOT laser beams for Rb and Yb   shares the same optical path through dichroic mirrors. The background vacuum pressure of the 2D-MOT chamber is consistently maintained below $10^{-9}$ mbar with a compact pump (Nextorr Z200).
    \begin{itemize}
    \item Rb: Zeeman slower beam power $P = 22$~mW, and frequency detuning $\Delta = -142$~MHz from the transition of $5S_{1/2}\ F=2 \rightarrow5P_{3/2}\ F'=3$ ($780.2$~nm). Each 2D-MOT beam power is $42$~mW with a diameter of $20$~mm and frequency detuning of $-31.8$~MHz (CF35 tubes).
    \item Yb: Zeeman slower beam power $P= 35$~mW, and frequency detuning $\Delta = -355$~MHz from the transition of $6s^2\ ^{1}S_{0}\rightarrow 6s6p\ ^{1}P_{1}$ ($398.9$~nm). Each 2D-MOT beam power is $30$~mW with a diameter of $20$~mm and frequency detuning of $-64$~MHz (CF35 tubes).
    \end{itemize}
    \item Magnetic field: Permanent magnets are employed to generate a gradient magnetic field for both the Zeeman slower and the 2D-MOT. The magnetic field profile is similar to the one described in \cite{PhysRevA.96.053415}. A gradually decreasing magnetic field slope is utilized for the Zeeman slowing, while a cross-zero gradient magnetic field is applied for the 2D-MOT. The maximum magnetic fields are $30$~G for Rb and $119$~G for Yb along the atom slowing direction, respectively. The magnetic field gradients for the Rb and Yb 2D-MOTs are set at $20$~G/cm. 
     \item Oven: Two separate ovens are used to generate atomic flux for Rb and Yb. Each oven has two independently temperature-controlled zones: one at the nozzle and the other at the body. The temperature of the oven nozzle is maintained approximately 20 degrees (Rb) or 30 degrees (Yb) higher than that of the body to prevent blockage in the capillary tubes.
     The capillary is $12$~mm long with a diameter of $0.45$~mm. The oven nozzle temperature is maintained at $110$~$^\circ$C for Rb and $410$~$^\circ$C for Yb, respectively. 
\end{itemize}
\begin{figure}
    \centering
    \includegraphics[width=\linewidth]{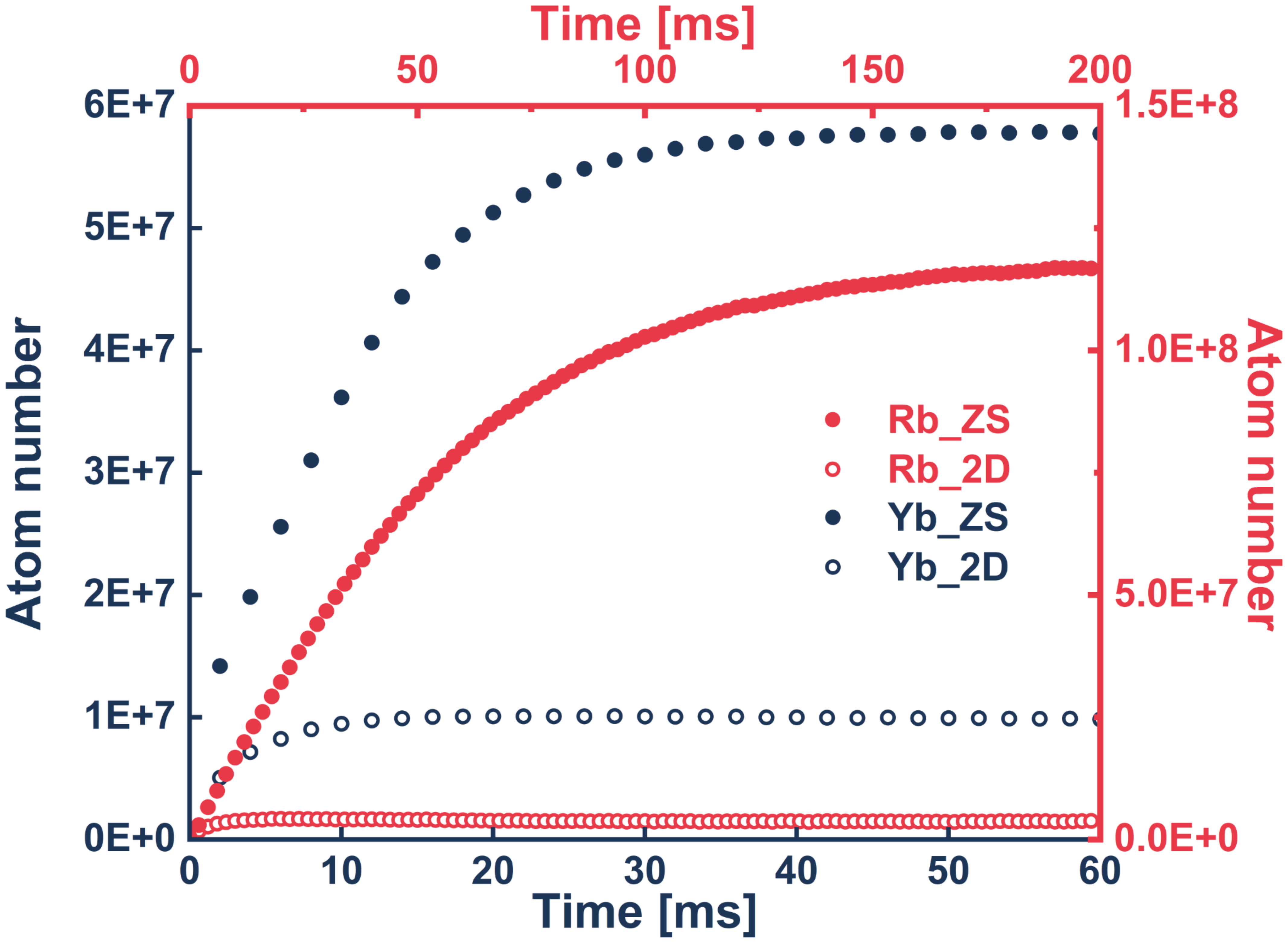}
    \caption{Enhanced time-dependent loading of $^{87}$Rb and $^{174}$Yb in 2D-MOTs utilizing Zeeman slowers.  The saturated atom numbers of Rb and Yb increase by a factor of 26.2, 5.7 in the presence of  the dual-beam Zeeman slower, respectively.  Atomic numbers are calculated by atomic fluorescence signals, which are captured every $2$~ms using a scientific CMOS camera for both Rb and Yb, respectively.}
    \label{fig7}
\end{figure}

 Figure~\ref{fig7} illustrates the time-dependent atom loading of $^{87}$Rb and $^{174}$Yb in the shared 2D-MOT, highlighting the significant enhancement achieved through the use of dual-beam Zeeman slowers for both species. For Rb, the atom number in the 2D-MOT reaches saturation around $200$~ms, achieving an equilibrium value of $1.1 \times 10^8$ with the Zeeman slower. In contrast, when only the 2D-MOT beam is used, the atom number is merely $4.2 \times 10^6$, indicating an enhancement factor of $26.2$. Similarly, for Yb, the equilibrium atom number in the 2D-MOT with the Zeeman slower is $5.7 \times 10^7$, compared to $1.0 \times 10^7$ with the 2D-MOT alone, resulting in an enhancement factor of $5.7$. These enhancement factors closely resemble those reported in Na (12$\times$)~\cite{10.1063/1.4808375} and Sr (4$\times$)~\cite{PhysRevA.96.053415} , where single Zeeman slower beams were employed.

A quantitative analysis of the loading curves, using exponential fitting to the model $N(t)=N_0[1-\exp(-t/\tau)]$ , reveals that the introduction of the dual-beam Zeeman slower increases the 2D-MOT loading rate to $1.2\times10^9$ atoms/s for Rb and $5.1\times10^{9}$ atoms/s for Yb. Moreover, the characteristic time constant (($\tau$) significantly rises from a few milliseconds to tens of milliseconds with the use of the dual-beam Zeeman slower. This finding is counterintuitive to what is reported in the literature~\cite{PhysRevA.102.013319}. We attribute this discrepancy to the short distance of ~$11.1$~cm between the oven nozzle and the 2D-MOT location in our compact setup. Atomic collisions, particularly those between ambient hot atoms and decelerated atoms, can alter the velocity distribution of atoms during the capture process, thereby requiring longer interaction times to achieve effective deceleration with the dual-beam Zeeman slower. Nonetheless, we observe that the equilibrium atom number in the 2D-MOT continues to increase steadily as the oven temperature rises. Notably, the background vacuum pressure in our 2D-MOT system remains below $10^{-9}$~mbar under high-temperature operation of the Rb and Yb, thanks to  differential pumping through capillary tubes.

To the best of our knowledge, the highest reported loading rate for $^{87}$Rb is $2\times 10^{10}$ atoms/s, achieved at a vacuum pressure of $4\times10^{-7}$~mbar~\cite{PhysRevA.74.023406}. It further demonstrated that the loading rate of Rb atoms increases linearly with the vacuum vapour pressure when it is below  $2\times10^{-7}$~mbar. Therefore, the estimated loading rate at $1\times10^{-9}$~mbar is about $5\times10^7$ atoms/s from this work. Another representative study, using a conventional $69$-cm single-beam Zeeman slower for $^{87}$Rb 3D-MOT loading, reported a loading rate of $3\times10^8$ atoms/s at an elevated oven nozzle temperature of $120$~degrees~\cite{PhysRevA.79.063631}. Compared with these studies, our system, employing a rather compact dual-beam Zeeman slower, yields substantial improvement of $^{87}$Rb atoms loading rate above $10^9$ atoms/s at a vacuum pressure below $10^{-9}$~mbar. 

The $^{174}$Yb MOT loading rate achieved in our setup is exceptionally notable. With a larger magnetic field gradient of $50$~G/cm, the MOT loading rate increases further to $8.0\times 10^{10}$ atoms/s (not shown in Fig.~\ref{fig7}). To the best of our knowledge, this is the first report to surpass $1\times 10^{10}$ atoms/s for Yb atoms at a moderate nozzle temperature of 410°C~\cite{PhysRevA.91.053405,PhysRevApplied.23.014004,Wodey_2021}. This underscores the advantage of a compact system for efficiently loading cold atoms, even those with high melting points~\cite{10.1063/5.0162128}. Angular dual-beam Zeeman configurations have been reported in previous studies for Dy~\cite{PhysRevA.101.063403}, Er~\cite{PhysRevA.102.013319}, and Yb~\cite{10.1063/5.0011361}. However, these designs are primarily intended to enhance the performance of conventional Zeeman slowers rather than operating independently.

It is important to note that, because the 2D-MOT chamber is shared for both Rb and Yb, the performance for each individual element is somewhat compromised. Consequently, there is significant potential for optimizing the performance for each element independently. Given that the atomic flux from the oven is highly directional, the background pressure in the 2D-MOT chamber could be further improved by incorporating an atomic beam shutter to dynamically control the hot atomic beam, turning it on and off as needed~\cite{10.1063/5.0123971}.
 
Our setup has been in operation for over a year, and no atomic deposits have been observed at optical windows for Zeeman slower beams, confirming the effectiveness of our method in reducing harmful atomic flux. This is essential for a compact Zeeman slower design.

\section{Discussion}\label{discussion}

To demonstrate the multifaceted capabilities of our Zeeman slower system, we present a comparative analysis of its performance across five distinct atomic species. Table \ref{table1} summarizes key metrics for each atom: maximal achievable velocity ($v_{\text{max}}$), final velocity threshold ($v_{\text{t}}$), most probable speed ($v_{\text{p}}$), effective cooling efficiency ($\eta_{\text{c}}$), and oven temperature (T). All simulations assume identical atomic vapor number densities $n$ in the oven to ensure consistent comparison conditions. The system geometry remains fixed across all cases: a Zeeman slower length of $L = 280$~mm, divergence angle $\theta_L = 3^\circ$, and beam diameter $d = 30$~mm. For $v_{\text{t}}$, we consider the presence of an associated 2D-MOT characterized by cooling beams with a diameter of $20$~mm and a saturation parameter $S_0 = 6$. Despite differences in atomic mass, transition properties, and thermal characteristics, approximately $21$\% of the flux from each species is transferred to the MOT region, underscoring robust adaptability of our design.

The results reveal notable disparities in performance across different species. Strikingly, ${}^{88}$Sr and ${}^{174}$Yb achieve cooling efficiencies ($\eta_{\text{c}} > 17$\%) nearly four times higher than that of ${}^{87}$Rb (5.5\%), despite their larger masses and broader thermal velocity distributions. This counterintuitive performance is attributed to the broad cycling transitions between $^1S_0$-$^1P_0$ states, which enhance photon scattering efficiency for these species. In contrast, ${}^{7}$Li exhibits reduced cooling effectiveness (3.2\%) due to its extremely high most probable speed ($v_p = 1197$ m/s), which requires higher oven temperatures (818 K) to achieve sufficient vapor pressure.

\begin{table}[h!]
\centering
\setlength{\tabcolsep}{9.5pt}
\begin{tabular}{
    >{}c        
    S[table-format=3.0]  
    S[table-format=3.0]
    S[table-format=4.0] 
    S[table-format=2.1] 
    S[table-format=4.0]  
}
\hline\hline 
\\[-1ex]
\multicolumn{1}{c}{\text{Atom}} & 
\multicolumn{1}{c}{$v_{\text{max}}$} & 
\multicolumn{1}{c}{$v_{\text{t}}$} & 
\multicolumn{1}{c}{$v_{\text{p}}$} &
\multicolumn{1}{c}{${\eta_{\text{c}}}$} & 
\multicolumn{1}{c}{${\text{T}}$} 
\\
& \multicolumn{1}{c}{(m/s)} & 
\multicolumn{1}{c}{(m/s)} & 
\multicolumn{1}{c}{(m/s)} & 
\multicolumn{1}{c}{(\%)} & 
\multicolumn{1}{c}{(K)} 
\\
\\[-1ex]
\hline
\\[-1ex]
$^{87}$Rb  & 203 & 30 & 260   & 5.5   & 353 \\
$^{174}$Yb & 432 & 65 & 272   & 17.8  & 752  \\
$^{88}$Sr  & 604 & 90 & 362   & 18.9  & 816 \\
$^{7}$Li   & 760 & 112 & 1197 & 3.2   & 818 \\
$^{168}$Er & 449  & 68 & 369  & 13.2  & 1534\\
\\[-1ex]
\hline\hline 
\end{tabular}
\caption{Performance metrics for different atomic species: maximal achievable velocity ($v_{\text{max}}$), final velocity threshold ($v_{\text{t}}$), most probable speed ($v_p$), effective cooling efficiency ($\eta_{\text{c}}$), and oven temperature (T). System parameters: $L=280$ mm, $\theta_L=3^\circ$, and Zeeman slower beam diameter $30$~mm. }
\label{table1}
\end{table}

To evaluate optimal Zeeman slower configurations, Table~\ref{table2} analyzes the length-dependent performance metrics using $^{87}$Rb as a reference. While increasing the slower length enhances maximal achievable velocity ($v_{\text{max}}$), this gain is accompanied by reduced transport efficiency ($\eta_{\text{t}}$). Intermediate-length systems exhibit optimal cooling performance: a $500$-mm configuration achieves peak effective cooling efficiency ($\eta_{\text{c}} = 6.7\%$ - nearly quadruple that of most compact $66$-mm designs at 1.7\%), while excessively long $700$-mm systems, though attaining the highest $v_{\text{max}}$ ($316$~m/s), suffer significant transport efficiency degradation ($\eta_{\text{t}} = 11.7$\%). This highlights the necessity of balancing deceleration capability with practical performance criteria.

Angled geometries  ($\theta_L =7^\circ$) provide strategic advantages in harmful flux management without substantial cooling efficiency penalties.  For instance, the 280-mm slower with angled geometry maintains nearly identical maximal achievable velocity ($v_{\text{max}}=201$~m/s vs.\ $203$~m/s for the standard design) while experiencing a $\sim18\%$ reduction in effective cooling efficiency ${\eta}_{\text{cap}}$ compared to non-angled systems of equal length. The enhancement factor reaches as high as 163 when compared to the scenario without Zeeman slower cooling (see Fig.~\ref{fig5}). The harmful atomic
flux in this angled design is significantly reduced to less than $0.01$\% (see Fig.~\ref{fig4}). Taking $L{'}$ into account, the total length of the angled 280-mm slower is only $\sim44$~cm. This design approach becomes particularly critical for high-melting-point species like $^{168}$Er (oven temperature $T=1534$~K), where elevated thermal conditions necessitate rigorous flux management without compromising essential cooling efficiencies. Please note that for the $280$-mm slower, the effective cooling efficiency $\eta_c$ increases  by $23$\% when the cooling beam size of the 2D-MOT is reduced to $10$~mm, compared to the non-angled version (data not shown in Table~\ref{table2}). This further confirms the anti-blooming effect discussed in Fig.~\ref{fig6}.

\begin{table}[h!]
\centering
\setlength{\tabcolsep}{10.5pt}
\begin{tabular}{
    >{}c
    S[table-format=3.0]
    S[table-format=1.1]
    S[table-format=2.1]
}
\hline\hline
\\[-1ex]
\multicolumn{1}{c}{\text{$L=2L_1$}} & 
\multicolumn{1}{c}{$v_{\text{max}}$} & 
\multicolumn{1}{c}{${\eta_{\text{c}}}$} & 
\multicolumn{1}{c}{${\eta_{\text{t}}}$} 
\\
\multicolumn{1}{c}{\text{(mm)}} & 
\multicolumn{1}{c}{\text{(m/s)}} & 
\multicolumn{1}{c}{(\%)} & 
\multicolumn{1}{c}{(\%)} \\
\\[-1ex]
\hline
\\[-1ex]
66         & 102     & 1.7 & 38.4 \\
111        & 130     & 3.0 & 33.9 \\
280(0$^\circ$) & 204 & 5.6 & 22.2 \\
280        & 203     & 5.5 & 22.3 \\
280(7$^\circ$) & 201 & 4.5 & 22.3 \\
500        & 269     & 6.7 & 15.2 \\
700        & 316     & 5.1 & 11.7 \\
\\[-1ex]
\hline\hline 
\end{tabular}
\caption{Maximal achievable velocity ($v_{\text{max}}$), effective cooling efficiency(${\eta_{\text{c}}}$) and transport efficiency (${\eta_{\text{t}}}$) for $^{87}$Rb with different lengths of Zeeman slowers. The oven temperature is fixed at $T=353$~K, and by default,  $\theta_L=3^\circ$ except special cases within parentheses. }
\label{table2}
\end{table}
  
\section{Conclusion}
Our dual-beam Zeeman slower design successfully mitigates the challenges posed by conventional single-beam systems, particularly the issue of residual atomic flux contaminating optical windows. By integrating two oblique laser beams with a capillary-array collimation system, we achieve efficient deceleration while significantly reducing harmful atomic flux and minimizing the spatial footprint of the apparatus. The demonstrated performance with rubidium and ytterbium, including a dramatic improvement in 2D-MOT loading efficiency and near-zero contamination levels, underscores the potential of this design for high-flux cold atom \cite{PhysRevA.108.023719}. Its compact geometry and scalability make it suitable for various atomic species, offering promising prospects for analogue and digital quantum computers built from neutral atoms. 

The miniaturization of cold-atom systems is essential for space-based quantum technologies, enabling precision metrology and gravitational studies in microgravity environments. Space station applications impose strict constraints on system size, power consumption, and robustness against mechanical vibrations~\cite{alonso2022cold,10.3389/fphy.2022.971059}. Our design might not only address these challenges but also ensure operational stability under launch stresses and long-term space conditions by incorporating optimized permanent magnets for passive field gradients. This advancement represents an important milestone in advancing cold-atom technology, providing a robust platform for reliable and portable quantum technologies with enhanced performance and reduced complexity.

\begin{acknowledgments}

We appreciate Mr. Huo's initial support in the MC simulation. This research was supported by the Guangdong-Hong Kong-Macau Greater Bay Area Quantum Science Center and funded through the Guangdong Province Quantum Science Strategic Initiative (Grant No.~GDZX2302005), National Natural Science Foundation of China (No.~12304348), Guangdong University Featured Innovation Program Project (2024KTSCX036), Guangzhou-HKUST(GZ) Joint Funding Program (No.~2025A03J3783), and Guangzhou Municipal Science and Technology Project (No.~2025A04J7077, No.~2025A03J3861, No.~2024A04J4351). J.F.C. acknowledges the support from NSFC No. 92265109, and the Guangdong projects under Grant No. 2022B1515020096. 
\end{acknowledgments}

\appendix

\section{Monte Carlo Simulation}
For cases where $L \ll L{'}$, we can simplify our analysis by adopting a solid angle model. In this scenario, it suffices to consider divergence along $\theta$ alone to estimate the proportion of atomic beam that exits the capillary array toward the optical window:
\begin{equation}
    P_{\text{harm}} = \iint_S f(\theta) \sin\theta \, d\theta \, d\phi,
\end{equation}
where $f(\theta)$ represents the angular distribution function under the given approximations (see Section~\ref{Transparent channel}), and $\theta$ and $\phi$ denote the polar and azimuthal angles, respectively.

However, when a longer $L$ is required for deceleration, the solid angle model becomes inadequate. To accurately estimate the proportion of harmful atomic flux reaching the optical window in our Zeeman slower system, we implemented a Monte Carlo (MC) simulation~\cite{PhysRevApplied.13.014013}. This method accounts for complex atomic trajectories, the influence of the capillary array’s structure, and parameters like the initial beam geometry and atomic velocity distribution.

The MC simulation incorporates a capillary array structure with an aperture diameter $D_{\text{ape}} = 0.45$ mm and length $L_{\text{ape}} = 12$ mm, yielding an aspect ratio $\beta = \frac{D_{\text{ape}}}{L_{\text{ape}}}$. We used established relationships to calculate the rubidium vapor pressure at 80°C, subsequently determining the mean free path ($\lambda_{\text{fp}}$) and Knudsen number ($\kappa_n$). These parameters are crucial in defining the angular flux distribution. Initial atomic positions and velocities are sampled from a Maxwell-Boltzmann distribution corresponding to the specified temperature. The angular distribution function is numerically determined to ensure that the emitted flux adheres to the correct divergence profile, as discussed in Section \ref{Transparent channel}.

To optimize computational efficiency, we impose reasonable cutoffs for the initial velocity components $v_x$, $v_y$, and $v_z$. Atomic trajectories are simulated using discrete time steps of $dt = 5$ $\mu$s, with position and velocity updates performed at each step according to:
\begin{align*}
    x(t+\Delta t) &= x(t) + v_x \cdot dt, \\
    y(t+\Delta t) &= y(t) + v_y \cdot dt, \\
    z(t+\Delta t) &= z(t) + v_z \cdot dt.
\end{align*}

 If an atom traverses the optical window region during its trajectory, it is classified as harmful flux. The MC simulation tracks the number of atoms that reach the optical window region, including those undergoing multiple collisions with the capillary walls. The proportion of harmful flux is quantified by:
\begin{equation}
    \frac{\Phi_{\text{harm}}}{\Phi_{\text{tol}}} = \frac{\text{number of harmful atoms}}{\text{total number of atoms}},
\end{equation}
where $\Phi_{\text{tol}}$ represents the total atomic flux.

The resulting MC simulation accurately predicts the harmful atomic flux, allowing for informed optimization of experimental parameters like capillary geometry, Zeeman slower length, and laser beam diameter.

\bibliography{apssamp}

\end{document}